\documentclass[12pt]{article}
\usepackage{epsfig}
\usepackage{amsmath}

\begin{document}


\begin{flushright}
\bf IFJPAN-V-05-05
\end{flushright}

\vspace{6mm}
\begin{center}
{\LARGE {\bf Upgrade of the Cellular General Purpose Monte Carlo Tool
             FOAM to version 2.06} }
\end{center}

\vspace{1mm}

\begin{center}
{\bf S.~Jadach}\\
\vspace{1mm}
{\em Institute of Nuclear Physics, Academy of Sciences,\\
  ul. Radzikowskiego 152, 31-342 Cracow, Poland,}\\
{\em and}\\
{\em CERN Department of Physics, Theory Division\\
CH-1211 Geneva 23, Switzerland}\\
\vspace{1mm}
{\em and}\\
{\bf P.~Sawicki} \\
\vspace{1mm}
{\em Institute of Nuclear Physics, Academy of Sciences,\\
  ul. Radzikowskiego 152, 31-342 Cracow, Poland}\\
\end{center}

\begin{abstract}
FOAM-2.06 is an upgraded version of FOAM, a general purpose, 
self-adapting Monte Carlo event generator. In comparison with
FOAM-2.05, it has two important improvements. New interface
to random numbers lets the user to choose from the three 
"state of the art" random number generators. 
Improved algorithms for simplical grid need less computer memory;
the problem of the prohibitively large memory allocation required for
the large number ($>10^6$) of simplical cells is now eliminated --
the new version can handle such cases even on the average desktop computers.
In addition, generation of the Monte Carlo events,
in case of large number of cells, may be even significantly faster.

\end{abstract}

\vspace{1mm}
\begin{center}
\em To be submitted to Computer Physics Communications
\end{center}

\vspace{13mm}
\noindent keywords: Monte Carlo (MC) simulation and generation, 
particle physics, phase space.

\vspace{15mm}
\begin{flushleft}
{\bf IFJPAN-V-05-05\\
     June~2005}
\end{flushleft}

\vspace{5mm}
\footnoterule
\noindent
{\footnotesize
$^{\star}$Supported in part by EU grant MTKD-CT-2004-510126,
          in the partnership with CERN PH/TH Division
}

\newpage

\section{Introduction}

{\tt FOAM} program is designed to be an universal MC event 
generator/integrator. It can be used for generating
weight one events and for evaluating integrals.
{\tt FOAM} employs combinations of two MC techniques: 
importance and stratified sampling known from older programs
oriented for these tasks VEGAS \cite{Lepage:1978sw}
and MISER \cite{Press:1989vk}. The most important and
most sophisticated part of the {\tt FOAM} algorithm
is the procedure of dividing the integration region into a system of 
cells referred to as a "foam". The cells in "foam" can 
be hyperrectangles, simplices or Cartesian product of both. 
The process of the cellular splittings is driven 
by the user defined distribution function such that minimization
of the ratio of the variance of the weight 
distribution to average weight (calculating integrals) 
or the ratio of maximum weight to the
average weight (generating events) is achieved.
For the detailed description 
of {\tt FOAM} algorithm we refer the interested reader to 
Refs.~\cite{Jadach:2002kn,Jadach:1999sf}

Until now {\tt FOAM} has passed practical tests in some 
applications to high energy physics.
We believe that foundations of the algorithm are
well established and our current work concentrates
on the updates of the program/algorithm toward better efficiency
and functionality.

At present, development path of {\tt FOAM} program goes in two directions.
In the recent paper ~\cite{mfoam:2005} we describe its new compact 
version -- {\tt mFOAM} -- 
with simplified use due to integration with ROOT system ~\cite{root:1997}
at the expense of slightly limited functionality.
In {\tt mFOAM} only hyperrectangular
foam is used. This seems to be sufficient for most probability distributions
encountered in the every-day practice.
However, in certain more advanced applications
the simplical grid may be definitely a better solution.
Hence, we do not abandon development of full version of 
{\tt FOAM}, which is intended for an experienced user,
in case when {\tt mFOAM} fails.

In the previous version of {\tt FOAM} simplical grid is encoded
in the program in the straightforward way: all vertices of all simplices
are stored directly in the computer memory.
This may lead to problem for large number of cells $N_c$ for
large number of dimensions $n$,
because amount of allocated memory grows roughly as a 
product of them, $\sim n\; N_c$.
In a complex problems with millions cells and $n>10$, 
the required amount of memory may exceed typical physical
memory size on a modern typical desktop computers. 

Let us note that this problem for the hyperrectangular foam
was already solved in the {\tt FOAM} development, by employing
sophisticated way of encoding system of cells in the computer memory.
With such a method, the memory consumption per hyperrectangle can be limited 
to a constant number per cell independent of $n$.
The main idea from this method is extended 
here to a simplical grid. As a byproduct we shall get also
a new faster and more stable method for computing volume of simplical cell,
without the need of determinant evaluation. 

In the presented new version 2.06 of {\tt FOAM},
the are two main improvements:
(i) the memory saving solution mentioned above,
(ii) the introduction of the new abstract base class -- {\tt TRND} -- 
which provides access to the library of
three high quality random number generators.
We do not follow in the development line 
of {\tt mFOAM} of introducing ROOT's objects whenever possible,
because {\tt FOAM} is meant to be optionally used as a stand-alone program,
without ROOT.
This is why we leave old organization of cells, with
the integer indices instead of pointers%
\footnote{This may look artificial; the reason for that is to 
   ensure persistency, when working with ROOT.}.

The paper is organized as follows: In Section 2 we discuss new 
memory saving solution for simplical grid. Next, in Section 3 
we describe changes in the code and compare performance of old
and new algorithms. 
Appendix covers some technical details relevant for efficient encoding
of the foam of simplical cells.

\section{Memory saving algorithms for simplices.}

\begin{figure}[htb]
\begin{center}
\psfig{file=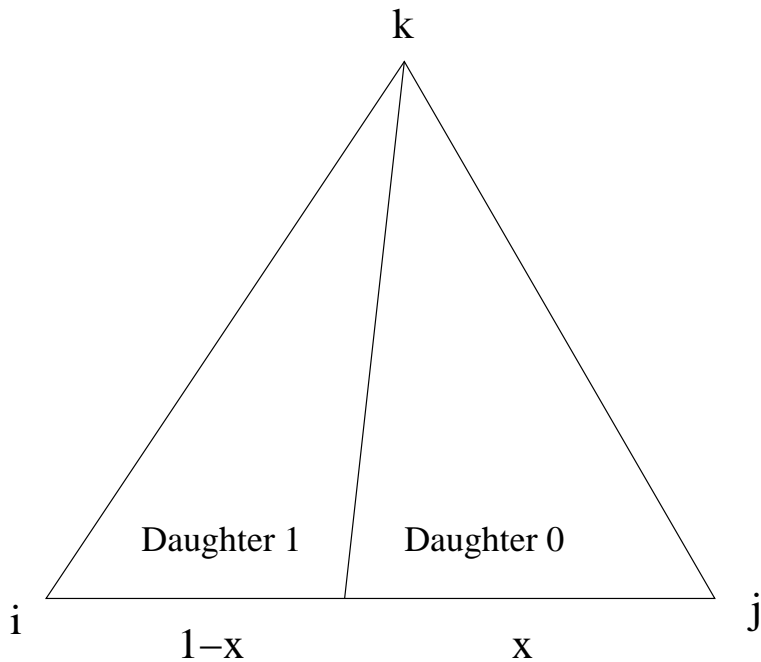}
\end{center}
\caption
{Split of a simplex.}
\label{split}
\end{figure}

In the algorithm (code) of {\tt FOAM}
simplices are organized as a linked tree structure.
Each cell (except the single root cell being the entire space)
has its parent cell.
All cells: parent and daughter cells are kept in the 
computer memory%
\footnote{That costs factor 2 overhead in the memory consumption.}.
Parent cells, which underwent division, we call ``inactive cells''
and the remaining are called ``active'' ones.
Each parent (inactive) cell has two daughters, active or inactive. 
The foam of cells is built during the exploration phase in
the process of a binary split of cells. 
Fig.~\ref{split} visualizes schematically the split of $n$-dimensional simplex. 
In the figure we show vertices, edges and division plane (line) on
the 2-dimensional plane spanned by the line of the ``division edge'',
which joins vertices $i$ and $j$, and the other vertex $k$, which
can be any of the remaining vertices, $k\neq i$, $k\neq j$.
The central line marks the intersection of our plane with the
division plane between two daughter cells.

Let us denote by $\vec{V}_i$ the Cartesian  {\em absolute coordinates}
of $i$-th vertex in the universal reference frame of the root cell.
In the {\tt FOAM} algorithm,
after the cell division a new vertex $V_{new} $ is put 
somewhere on the division edge $(i,j)$ of the simplex
\begin{equation}
\vec{V}_{new} = x \vec{V}_i + (1- x) \vec{V}_j.
\end{equation}
The geometry of the division is defined unambiguously by specifying 
the division edge (two integers) and the division parameter $x$
(one number of {\tt double} type).

In the original {\tt FOAM} absolute positions of all vertices $\vec{V}_i$
are recorded in the computer memory as a list of Cartesian vectors 
defined in a universal reference frame tied up to the root cell. 
The above vector algebra is done in terms od such vectors, all in the same
universal reference frame. 
(This is why memory consumptions grows with the dimension.)

On the other hand, during the MC exploration of each cell, 
one generates randomly MC points with the uniform distribution within
any given cell using another kind of the internal coordinates%
\footnote{For the internal coordinates $\lambda_i, i=1,2,\ldots ,n+1$
 we keep convention that $\lambda_i \in \langle 0,1 \rangle $
 and $\sum \lambda_i\leq 1$. Furthermore one of them
 is equal zero $\lambda_k=0$. See also Appendix.}
specific for a given simplex cell,
which cannot be used for any other cell than this one.
There is, however, very important exception from the above restriction:
we may establish connection between the internal variables
of the parent cell and the two daughter cells!
This opens the way of eliminating the need of storing Cartesian vectors
of all vertices $\vec{V}_i$, and thus reducing the memory consumption.

The basic idea is to look up recursively the linked tree of ancestors,
starting from a given active cell and finishing with the root cell.
In a single step upward one translates the internal coordinates 
of the daughter cell into internal coordinates of the parent cell.
In the end the internal coordinates of any point in any active cell
gets translated into the internal coordinates
of the root cell, which are universal!
In this way, we only need to know the $n+1$ Cartesian vectors
of the vertices of  the root cell,
in order to translate every point in every active cell
to an universal Cartesian vector.
The universal Cartesian vector is, of course, needed to evaluate
the distribution function
(which knows nothing about foam of cells and its internal coordinates),
and is also used for other purposes
by the external user of the {\tt FOAM} object.

The above translation from the daughter to the parent internal variables
is more complicated for the simplical cells than for hyperrectangular cells
and we shall describe it in the following in more details.

The essential point in the proper connecting of the internal coordinates
of the parent and daughter cell
is the careful choice of the origin point of the coordinate 
system of the internal variables in the relevant cells.
The definition of the internal coordinates
always distinguishes one of the vertices of a given cell%
\footnote{The one for which $\lambda_k=0$.}.
Let us call this vertex the {\em origin vertex}.

As already indicated in Fig.~\ref{split}
the division point is located on the interval $(i,j)$.
Following notation in the figure,
if daughter number 0 has the origin vertex being vertex $j$
then the transformation of the internal coordinates between the daughter cell 
($\lambda_l$) and the parent cell ($\lambda'_l$)
is just the following dilatation of just one $i$-th coordinate
\begin{equation}
\label{l1} 
  \lambda^{'}_i= x \lambda_i.
\end{equation}
Similarly, for daughter number 1 and parent cell sharing
the origin vertex $i$ we shall have 
\begin{equation}
\label{l2}
  \lambda^{'}_j = (1-x) \lambda_j.
\end{equation}
What to do in a typical situation when the parent and daughter cell
do not share the origin vertex as described above?
The solution is simple, change origin vertex to the vertex which 
fits the dilatation scheme outlined above.
The resulting additional transformation of the internal coordinates
is described in Appendix.

The entire algorithm of building the absolute
Cartesian coordinates using information stored
in the ancestry line of the active cell can be summarized as follows:
\begin{enumerate}
\item[(a)] 
  For a given current active cell look for its parent.
  If our cell is already the root cell then 
  translate $\lambda$'s into Cartesian coordinates and {\em exit}, 
  else go to point (b).
\item[(b)]
  Check if the origin vertex is in convenient location and
  optionally redefine the origin vertex
  performing transformation of Eq.~(~\ref{shift}) in Appendix.
  Go to point (c).
\item[(c)] Perform the transformation of Eq.~(\ref{l1}) or (\ref{l2}).
  Replace the current cell by its parent cell and return to point (a).
\end{enumerate}

From now on it is obvious that we do not need to know and store
the Cartesian vectors for all vertices of the simplical cells.
However, for establishing the proper total normalization of
the {\tt FOAM} integrand over each cell we also need to know the volume
of the cell.
In the original {\tt FOAM} algorithm it was evaluated using determinant
of the Cartesian vectors of the vertices.
Of course, with some effort, using the transformations described
above we could calculate Cartesian vectors of all $n+1$ vertices
and plug them into determinant.
Luckily, it is not necessary, because
we have found an alternative very simple method of calculating
the volume of the simplical cell in which the information
in the ancestry line of a cell is used, similarly as in
decoding of absolute coordinates described above.
Again, climb up the linked tree toward the root cell
and at each step we perform the transformation relating volume of 
the cell $V$ and its parent $V'$, which is very simple.
Following notation of Fig.~\ref{split},
for daughter number 0 the following rule applies
\begin{equation}
V= x V'
\end{equation}
while for daughter number 1 we should apply
\begin{equation}
V= (1-x) V'.
\end{equation}
The above method is fast and it also turns out to be more stable numerically
than the usual method based on evaluation of the determinants,
especially  at large $n$.

\section{Changes in code} 
In the following we shall summarize on the changes in the C++ source
code which implement the new method of coding simplical cells
and other changes.

\subsection{TFOAM and TFCELL class}
The algorithm described above for ``in flight'' decoding of
the absolute coordinates of the MC points inside any cell
(without the need of knowing the vertex positions) and algorithm for
volume calculation are implemented as two additional methods
in {\tt TFCELL} class: sophisticated 
translator of coordinates: {\tt GetXSimp(TFVECT \&, TFVECT \& , int )} 
and volume cell calculator {\tt CalcVolume(void)}. 
The method {\tt GetXSimp} is used in few places in program to provide
argument for the distribution function and during event generation
to translate internal coordinates of event to Cartesian coordinates, 
required by the user.  
The method {\tt  CalcVolume} calculates the volume of the cell. 
It replaced permanently  the old method {\tt  MakeVolume }
based on determinants.

The new solution with ``in flight'' transformation of coordinates is active in
the default configuration.
However, it might be sometimes convenient to switch back
to the old algorithm. For instance it is required if one
wants to use the plotting procedures: {\tt  LaTexPlot2dim } or
{\tt  RootPlot2dim }. The same applies for certain debugging methods.
To address these issues we have introduced in the {\tt TFOAM} class a new
switch {\tt m\_OptVert } and corresponding setter method
{\tt SetOptVert( int ) }. This switch has an analogous meaning to
{\tt m\_OptMCell} switch already introduced for hyperrectangles.
In order to activate storing positions of vertices in the memory
the user may invoke {\tt FoamX->SetOptVert(0) },
which resets {\tt m\_OptVert } to 0 (the default value is 1)
before the initialization of the {\tt FOAM} object (MC generator).

\begin{table}[hbt]
\centering
\begin{small}
\begin{tabular}{|c|p{2cm}|p{2cm}|p{2cm}|}
\hline
$N_c$      & $2.5 \times 10^5$ &  $5 \times 10^5$ & $10^6$     \\
\hline
 OptVert=  & \multicolumn{3}{c|}{ 0}                            \\
\hline
     Mem (Mby)  & 143                &  211           & 346    \\
    CPU (min)  &  118               &  392           & 1453  \\
\hline
 OptVert=   & \multicolumn{3}{c|}{ 1}                           \\
\hline
     Mem (Mby)  & 116                &  156             & 243     \\
     CPU (min)  & 67             & 156                  & 401 \\
\hline
\end{tabular}
\end{small}
\caption{\sf Performance comparison of new and old algorithms.}
  \label{tab:Performance}
\end{table}

In order to compare CPU and memory consumption in new and old 
algorithms  we performed some tests for function {\tt Camel} 
included in   {\tt TFDISTR } class. In each test we built grid 
consisting of  a few hundred  thousands to million 6-dimensional 
simplical cells  and generate  200k events. For computations we 
used high-end PC  equipped with AMD64 processor running with 2 GHz 
clock and 2 Gby RAM. In Tab.~\ref{tab:Performance} we compare total 
virtual memory consumption, shown  by {\tt top} utility program and 
CPU time needed to  complete each task. 

The memory consumption is significantly lower for new algorithms.
Moreover difference in memory consumption between old and new 
algorithms scales linearly, as expected, with the number of cells.
For the finest grid studied reaches above 100 Mby 
\footnote{In the same test repeated on SGI-2800 supercomputer
we found even deeper reduction. However we are in serious doubt
if the memory usage shown by the queue system is correct.}.
Obviously for the larger number $N_c$ memory consumption in old 
algorithms quickly becomes critical: program with old algorithms
can not be run at all while computations for new algorithms are
still feasible. To see this barrier we repeated the last test with  
$N_c=10^6$ on older desktop computer with 500 Mby of RAM. We succeeded
to complete it only with new algorithms. Program switched to old
algorithms did not deliver the result even after a week probably
due to heavy swap. 

As we see, new algorithms are also significantly faster. At 
this moment  we can not give detailed explanation to this rather 
surprising observation. Definitely some role can play fact that 
vertices in old  algorithms are accessed in random pattern. Rarely 
used vertices have tendency to be moved outside the cache memory 
or if the foam has indeed many cells even outside physical 
RAM to swap disk. This results in many cache conflicts or even worser
swapped vertices have to be read from disk. These operations introduce
significant time overhead which can slow down the program  
progress more than  additional modest computational effort in 
new algorithms.  

From the above tests we conclude that our memory saving solution was 
worth an extra programming effort and indeed improved 
performance of {\tt FOAM} with simplical cells.

\subsection{TRND - collection of random number generators}
\label{TRND}

The essential new auxiliary part of the {\tt FOAM} package is a small library
of the three random number generators inheriting from the pure Abstract
Base Class (interface) {\tt TRND}.
In its construction we had the following very clear specification in mind:
\begin{itemize}
\item The library of the random number generators should be {\em extensible},
  i.e. its user could easily add another random number generator
  with well defined minimum of functionality. 
  This implies that all r.n.g. classes should inherit from
  certain relatively simple pure abstract base class
  (interface).
\item The minimum functionality is defined as follows:
  \begin{itemize}
    \item Possibility to set initial ``seed'' in form of just one integer.
    \item Existence of a method to generate a single uniform random number.
    \item Existence a method to generate a series of the uniform random 
          numbers in a single call.
    \item It should be easy to record status of random number generator
          and restart it using recorded data. This is, of course, assured
          by the persistency mechanism of ROOT.
  \end{itemize}
\item The library of the random number generators should be {\em versatile}
  i.e. it should include at least one example of the generator,
  which features the best possible randomness (RANLUX), 
  one which is fast and has extremely long series (Mersenne Twister),
  and possibly some legacy code,
  which has been widely used in the past (RANMAR).
  From these options user may easily choose something which fits his particular
  Monte Carlo project.
\end{itemize}
After reviewing the existing libraries of the r.n. generators 
we concluded that none of them%
\footnote{The class {\tt TRandom} in ROOT almost fulfills our specification.
  We want, however, that {\tt FOAM} is able to work also
  in the stand-alone mode without ROOT.
  In addition {\tt TRandom} has a lot of 
  extra functionality which we do not need
  and it still does not include RANLUX.}
does meet the above specification,
and we decided to create such a library of our own,
with our own universal and minimalistic {\em interface}.
The user of {\tt  FOAM} may easily add his own favorite 
r.n. generator to this collection.

Abstract Base Class {\tt TRND} provides user interface to collection
of the random numbers generators.
It replaces the old class {\tt TPSEMAR}
of the original {\tt FOAM}. At present,
three random number generators are included: {\tt TRanmarEngine}
is an implementation of the well known RANMAR generator
\cite{Marsaglia:1990ig}.
Class {\tt TRanmarEngine} is practically identical with
the old {\tt TPSEMAR} class and both produce the same sequences
of the MC events, if initialized with the same seed.
We recommend it as a default one because it is fast and reliable.

RANLUX is the best of known available random number generators
in the literature~\cite{Luscher:1993dy}.
It is implemented here under the name
{\tt TRanluxEngine}, following closely ref.~\cite{James:1993vv}.
This generator has excellent spectral properties and user
defined ``luxury level''.  At the highest possible level=4
its 24 bits of mantissa are completely chaotic,
however, already for default level=3 its quality is high enough
for most of practical applications.

The third choice included in our small
library of the random number generators is {\tt TRMersenneTwister}
- implementation of the recently developed Mersenne Twister random
number generator, which is very fast and has huge period
$2^{19937}-1$ \cite{mtwistor}.
As an example: the following line of code
\begin{verbatim}
TRND *PseRan   = new TRanmarEngine();  // Create random number generator
\end{verbatim}
creates an instance of RANMAR generator with the default value of the
so-called seed.

Most of the code and method names were adopted from CLHEP
class of random numbers generators \cite{clhep}. All generators
are fully persistent; their state can be written into disk
and restored later using persistency mechanism of ROOT.

The great profit from the persistency implementation
is that the class objects (random number generators) restored
from the disk do not need any reinitialization -- they
can be just used immediately to generate
the next random number, following the one generated before the disk-write,
as if there was no disk-write and disk-read at all!

\section{Conclusions}
{\tt FOAM} package has reached  the level of the mature MC tool
to be used by the advanced builders of the MC event generators.
In this work we implement certain technical improvements in the code,
without any essential modification of the MC algorithm.
The main change in the presented new version 2.06 is the new
more efficient method of encoding the simplical foam of cells
in the computer memory.
The sophistication of the simplical foam is now upgraded
to the same level as of the hyperrectangular one. 
In addition, we add a more sophisticated library of the random number
generators with the universal interface.
The other features of the package remains the same as in version 2.05.

\section*{Acknowledgments}
The work was partially supported by EU grant MTKD-CT-2004-510126,
realized in the partnership with CERN PH/TH Division.
We thank to ACK Cyfronet AGH Computer Center
for granting us access to their PC clusters and supercomputers
funded by the European Commission grant
IST-2001-32243 
and the Polish State Committee for Scientific Research grants:
620/E-77/SPB/5PR UE/DZ 224/2002-2004,
112/E-356/SPB/5PR UE/DZ 224/2002-2004, 
MNiI/SGI2800/IFJ/009/2004 and
MNiI/HP\_K460-XP/IFJ/009/2004.

\vfill\newpage\noindent
{\bf\Large Appendix}
\vspace{2mm}

A simplex in $n$ dimension has $n+1$ vertices and each of them can 
be regarded as an origin point of coordinate system. 
The other $n$ vertices define vector directions which span
the vector space. More precisely, arbitrary vector $\vec{ P}_i$,
connecting i-th vertex with a certain point $P$ can be written as 
\begin{equation}
\vec{ P}_i= \sum_k \lambda_k \vec{ w}_k  \; \; \;  (k \neq i)
\end{equation}
in the internal basis of vectors $\vec{ w}_k=\vec{ V}_k -\vec{ V}_i$,
where $\vec{ V}_k$ is absolute position of k-th vertex. There 
are $n$ internal coordinates because $\vec{ w}_i=0$. However, it is 
convenient to adopt convention that there are $n+1$ coordinates 
and $\lambda_i=0$.

In our algorithm we shall use coordinates $(\lambda_l)$  to parametrize 
position of points {\em inside } the simplex or eventually on its surface. 
This leads to additional conditions
\begin{eqnarray*}
\lambda_k & \geq  & 0  \; \; \;  (k \neq i) \\
\sum_{k \neq i} \lambda_k  & \leq & 1. 
\end{eqnarray*} 


\begin{figure}[htb]
\begin{center}
\psfig{file=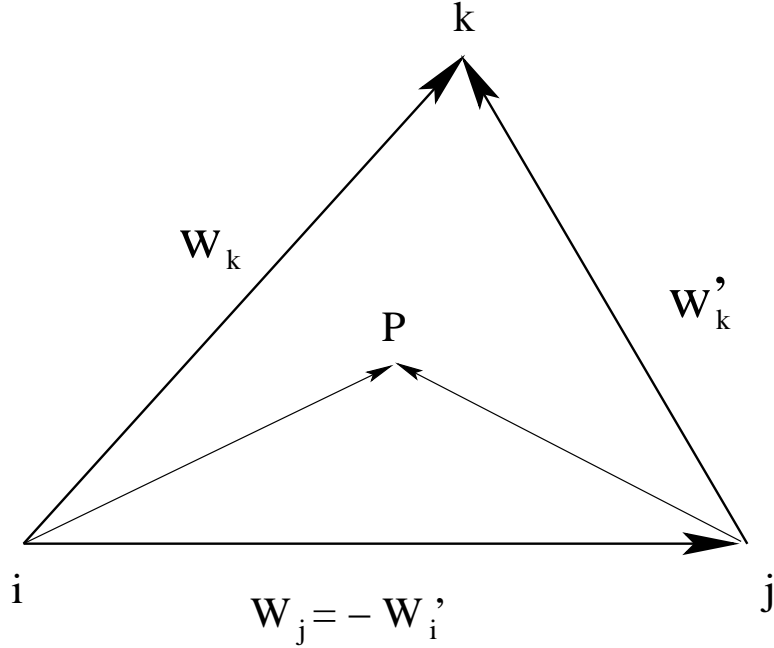}
\end{center}
\caption
{Transformation of internal coordinates under change of origin point. }
\label{vectors}
\end{figure}

An important component of our memory saving algorithm is
transformation that changes the origin point of coordinate
system from one vertex to another. In order to derive 
transformation law one has to relate coordinates of 
point $P$, residing inside the simplex, in the old and new 
basis (see Fig.~\ref{vectors}). For the vector $\vec{ P}_i$ 
originating from i-th vertex we have
\begin{equation}
\vec{ P}_i  =  \sum_{k \neq i} \lambda_k \vec{ w}_k
\end{equation}
and similarly for vector $\vec{ P}_j$ originating from j-th vertex
\begin{equation}
\vec{ P}_j  =   \sum_{k \neq i} \lambda_k \vec{ w}_k -\vec{ w}_j.
\end{equation}
Since $\vec{ w}_i^{'}= -\vec{ w}_j$ and 
$ \vec{ w}_k= \vec{ w}_k^{'}-\vec{ w}_i^{'}$
we have further
\begin{equation}
\vec{ P}_j = \sum_{k \neq i} \lambda_k (\vec{ w}_k^{'}-\vec{ w}_i^{'}) 
+ \vec{ w}_i^{'}.
\end{equation}
Finally
\begin{equation}
\vec{ P}_j = \sum_{k \neq i, k \neq j} \lambda_k \vec{ w}_k^{'} +
(1- \sum_{k \neq i}\lambda_k ) \vec{ w}_i^{'}.
\end{equation}
Coefficients before vectors $\vec{ w}^{'}$  are now the new internal 
coordinates. The transformation law reads as follows
\begin{equation}
\begin{split}
\lambda_k^{'} & = \lambda_k \; \; \; (k \neq i, k \neq j) \\
\lambda_i^{'} & = 1- \sum_{k \neq i} \lambda_k  \label{shift}\\
\lambda_j^{'} &  =0.
\end{split}
\end{equation}

\end{document}